\documentclass[runningheads]{llncs}
\usepackage[T1]{fontenc}
\usepackage{booktabs}
\usepackage{amsmath}
\usepackage{graphicx}
\usepackage{cite}
\usepackage{url}
\usepackage{amsmath, amssymb} 
\usepackage{algorithm} 
\usepackage{algpseudocode} 
\usepackage{mathtools} 
\usepackage{microtype}             
\usepackage{comment}

\begin{document}
\title{FlashMLA-ETAP: Efficient Transpose Attention Pipeline for Accelerating MLA Inference on NVIDIA H20 GPUs}
\titlerunning{FlashMLA-ETAP: Efficient Transpose Attention Pipeline}
%
%
\author{Pengcuo Dege\inst{1} \and
Qiuming Luo\inst{2} \and
Rui Mao\inst{2} \and
Chang Kong\inst{3}\thanks{Corresponding author}
}
\authorrunning{P. Dege et al.}
%
\institute{Tencent, Beijing, China \\
\email{pengcuoze@tencent.com} \and
College of Computer Science and Software Engineering, Shenzhen University, Shenzhen, China\\
\email{\{lqm,mao\}@szu.edu.cn} \and
College of Artificial Intelligence, Shenzhen Polytechnic University, Shenzhen, China\\
\email{kongchang@szpu.edu.cn}
}
\maketitle              
\begin{abstract}
Efficient inference of Multi-Head Latent Attention (MLA) is challenged by deploying the DeepSeek-R1 671B model on a single Multi-GPU server. This paper introduces FlashMLA-ETAP, a novel framework that enhances MLA inference for the single-instance deployment scenario on NVIDIA H20 GPUs. We propose the Efficient Transpose Attention Pipeline (ETAP), which reconfigures attention computation through transposition to align the KV context length with the \(M\)-dimension in WGMMA operations, significantly reducing redundant computations. FlashMLA-ETAP achieves a 2.78x speedup over FlashMLA at 64K sequence length (batch size 16), with 5.24x and 4.94x improvements over FlashAttention-3 and FlashInfer, respectively, while maintaining numerical stability with a 15.2x lower RMSE (\(1.25 \times 10^{-5}\)) than FlashAttention-3. Furthermore, ETAP's design enables seamless integration into frameworks like FlashAttention-3 and FlashInfer, supported by a detailed theoretical analysis. Our work addresses a critical gap in resource-constrained inference, offering a scalable solution for mid-tier GPUs and paving the way for broader adoption in hardware-aware optimization. Code is available at \url{https://github.com/pengcuo/FlashMLA-ETAP}.

\end{abstract}
\section{Introduction}
The Transformer architecture \cite{vaswani2017} has become a cornerstone of modern artificial intelligence, driving significant advances in natural language processing \cite{devlin2019}, computer vision \cite{alexey2020}, and multi-modal learning \cite{radford2021}. Central to this architecture is the attention mechanism, with variants such as Multi-Head Attention (MHA) and Multi-Head Latent Attention (MLA) \cite{deepseek-ai2024} enabling models to capture long-range dependencies across sequences. These mechanisms are critical for applications like large-scale language model decoding, video understanding, and real-time multi-modal inference. However, the quadratic computational complexity of attention operations, scaling with the square of the sequence length, poses a substantial bottleneck, particularly during inference for long-context tasks. This challenge is especially pronounced on mid-tier GPUs like the NVIDIA H20, which, with a modest 148 TFLOPS FP16 compute capability compared to the 1979 TFLOPS of high-end GPUs like the H100 or H800, struggles to achieve efficient hardware utilization. Architectural constraints, such as the padding requirements of WarpGroup Matrix-Multiply-Accumulate (WGMMA) instructions, further exacerbate this inefficiency when deploying the DeepSeek-R1 671B model \cite{deepseek-ai2025r1} on a single 8-GPU H20 server, where 128 heads are split into 16 per GPU, causing redundant padding as the \(M\)-dimension (16) falls below WGMMA’s minimum of 64, often reducing compute utilization to below 25\% during decoding phases with short queries and long key-value (KV) contexts in this single-instance deployment scenario \cite{deepseek-ai2025}.

Significant research has sought to mitigate these challenges, with a range of approaches targeting both algorithmic and hardware-level optimizations \cite{dao2022,dao2023,shah2024,ji2025,flashmla2025,ye2025}. Early efforts, such as sparse attention and low-rank approximations, aimed to reduce complexity by limiting the scope of attention or approximating matrices, though these methods often compromise model expressiveness, particularly for tasks requiring fine-grained context \cite{huang2024}. More recently, the FlashAttention series, starting with FlashAttention \cite{dao2022} and evolving through FlashAttention-2 \cite{dao2023} and FlashAttention-3 \cite{shah2024}, introduced I/O-aware tiling, online softmax computation, and asynchronous execution, leveraging high-end GPU capabilities like Tensor Memory Accelerator (TMA) and low-precision Tensor Cores. Building on FlashAttention-3, DeepSeek’s FlashMLA framework \cite{flashmla2025} tailored these optimizations for MLA, incorporating low-rank joint compression to minimize key-value (KV) cache sizes during inference, a critical factor for large models like DeepSeek-V3 and DeepSeek-R1. Despite these innovations, a significant gap persists: frameworks like FlashAttention-3 and FlashMLA, while highly effective on high-end GPUs like the NVIDIA H100 and H800, underutilize the constrained compute capabilities of mid-tier GPUs like the H20. This underutilization stems from inefficiencies such as redundant padding in WGMMA instructions, caused by insufficient memory necessitating head splitting, reducing the per-GPU head count below the WGMMA minimum \(M\)-dimension, and suboptimal memory access patterns, particularly in decoding scenarios with long KV contexts, highlighting the need for hardware-specific optimizations to enable efficient inference in this single-instance deployment scenario.

To address this critical gap, we propose FlashMLA-ETAP, an advanced iteration of the FlashMLA framework, specifically engineered to enhance the efficiency of MLA inference on mid-tier GPUs such as the NVIDIA H20. At the heart of our approach lies the Efficient Transpose Attention Pipeline (ETAP), a novel computation mode designed to overcome the inherent inefficiencies of existing attention mechanisms on hardware with limited computational resources. ETAP reconfigures the traditional attention computation process by introducing strategic transposition, a technique that reorients the alignment of key computational dimensions to mitigate the redundant padding overhead imposed by WGMMA instructions, caused by insufficient memory necessitating head splitting, reducing the per-GPU head count below the WGMMA minimum \(M\)-dimension, on the H20’s constrained 148 TFLOPS FP16 compute capability. This reorientation specifically targets the asymmetry prevalent in decoding scenarios, where short query lengths—often limited to a handful of tokens—contrast sharply with long key-value (KV) context lengths, which can extend to tens of thousands of tokens. By aligning the KV context length with the $M$-dimension in WGMMA operations, ETAP minimizes redundant computations, thereby optimizing the utilization of the H20’s modest compute resources and reducing the memory access bottlenecks that plague conventional approaches.

The theoretical foundation of ETAP rests on exploiting the dimensional asymmetry between query and KV context lengths during inference, a characteristic that is particularly pronounced in autoregressive tasks. This approach leverages the inherent structure of attention operations, redefining the computation flow to prioritize efficiency over the traditional, query-centric processing paradigm. The transposition strategy not only alleviates the redundant padding requirements that diminish compute efficiency—often dropping below 25\% on H20 GPUs—but also enhances the scalability of the attention mechanism across varying context lengths. To realize these benefits, ETAP is seamlessly integrated into the FlashMLA framework, which already incorporates low-rank joint compression to manage KV cache sizes effectively. This integration preserves FlashMLA’s strengths while amplifying its performance through ETAP’s optimized computation mode, resulting in a framework that achieves significant improvements in inference efficiency and numerical stability. Furthermore, the design of ETAP is intentionally generalizable, suggesting its potential to be adapted to other attention frameworks, such as FlashAttention-3 and FlashInfer \cite{ye2025}, thereby offering a versatile solution that could catalyze the development of hardware-aware inference optimizations across a broader range of GPU architectures. Through this work, we aim to lay a robust foundation for democratizing efficient inference on mid-tier GPUs, addressing a pressing need in resource-constrained environments and opening new avenues for future research in computational efficiency.

In summary, this paper makes the following key contributions:
\begin{itemize}
    \item We introduce the Efficient Transpose Attention Pipeline (ETAP), a novel computation mode that leverages transposition to reduce redundant calculations, tailored to the architectural constraints of mid-tier GPUs like the H20.
    \item We develop FlashMLA-ETAP, enhancing FlashMLA with ETAP to enable efficient MLA inference on mid-tier GPUs, with ETAP designed to be easily integrated into other attention frameworks such as FlashInfer and FlashAttention-3.
    \item Our experiments show that the FlashMLA-ETAP framework achieves impressive performance on the NVIDIA H20 GPU, delivering up to a 2.78x speedup over the baseline FlashMLA framework.
\end{itemize}

\section{Related Work}
The pursuit of efficient attention mechanisms has been a central focus in deep learning research, driven by the increasing scale of Transformer-based models and their application to long-sequence tasks \cite{devlin2019,alexey2020,radford2021,kwon2023,sglang2024,deepseek-ai2024,deepseek-ai2025,deepseek-ai2025r1}. This section reviews key developments in attention optimization, emphasizing their relevance to inference on mid-tier GPUs, with particular attention to the challenges posed by hardware like the NVIDIA H20, which is limited by its 148 TFLOPS FP16 compute capability.

Early efforts to mitigate the quadratic complexity of attention focused on algorithmic approximations. Sparse attention mechanisms, such as those in the Sparse Transformer \cite{child2019} and Longformer \cite{beltagy2020}, reduce computational cost by attending to a subset of tokens, achieving linear or near-linear complexity. Similarly, low-rank approximations like Linformer and Performer project attention matrices into lower-dimensional spaces, trading off expressiveness for efficiency. While these methods enhance scalability for specific tasks, they often compromise the rich contextual understanding required for fine-grained dependencies, limiting their suitability for long-context inference scenarios where full attention is beneficial.

A parallel thread of research has targeted hardware-aware optimizations to maximize GPU efficiency. The FlashAttention framework \cite{dao2022} pioneered I/O-aware tiling and online softmax computation, reducing high-bandwidth memory (HBM) accesses to improve performance on modern GPUs. FlashAttention-2 \cite{dao2023} extended this with sequence parallelism and work partitioning, while FlashAttention-3 \cite{shah2024} further refined the approach by leveraging advanced features like Tensor Memory Accelerator (TMA), optimizing for high-end GPUs such as the H100 and H800. DeepSeek’s FlashMLA \cite{flashmla2025}, an open-source adaptation of FlashAttention-3 available on GitHub, tailors these techniques for MLA, employing low-rank joint compression to minimize key-value cache sizes during inference—a crucial optimization for large models like DeepSeek-V3 \cite{deepseek-ai2025}. However, these frameworks, designed primarily for high-end GPUs, encounter significant inefficiencies on mid-tier hardware like the H20 in single-instance deployment scenarios. Their reliance on advanced features and tiling strategies often leads to redundant padding, caused by insufficient memory necessitating head splitting, reducing the per-GPU head count below the WGMMA minimum \(M\)-dimension, underutilizing the H20’s 148 TFLOPS FP16 compute capability.

Beyond these core approaches, alternative strategies have emerged to address attention efficiency. Distributed attention mechanisms, such as Ring Attention \cite{liu2023}, partition computation across multiple devices to handle extremely long sequences, but they assume access to multi-GPU setups, which may not be viable in resource-constrained environments. Hybrid architectures like Mamba \cite{gu2024} and RWKV \cite{peng2023} combine attention with state-space models to balance efficiency and expressiveness, yet their divergence from standard Transformer designs limits their direct applicability to MLA.

The increasing need for efficient inference on mid-tier GPUs in resource-constrained environments such as academic labs highlights the limitations of existing approaches. Collectively, these methods fail to fully address the unique challenges of mid-tier hardware, where constrained compute power and redundant padding inefficiencies hinder the deployment of advanced attention models. To tackle this gap, we propose a novel computation mode that optimizes inference efficiency on mid-tier GPUs, as detailed in the subsequent chapters.

\section{Methodology}
The limitations of existing attention optimization frameworks, as highlighted in the prior section, underscore a critical challenge: while approaches like FlashAttention-3 and FlashMLA achieve impressive efficiency on high-end GPUs, they fall short on mid-tier hardware such as the NVIDIA H20 due to its constrained 148 TFLOPS FP16 compute capability. Our analysis reveals that FlashMLA, despite its low-rank compression for MLA inference, suffers from low compute utilization—often below 25\%—stemming from redundant padding in WGMMA instructions and inefficient memory access patterns. These inefficiencies are particularly pronounced during inference, where long KV context lengths and short query lengths exacerbate the mismatch with the H20’s architecture. Motivated by this gap, we propose a novel solution that reconfigures the attention computation process to align with the H20’s capabilities, laying the foundation for significant performance improvements. This chapter introduces the Efficient Transpose Attention Pipeline (ETAP), a new computation mode designed to reduce redundant calculations, and details its integration into FlashMLA to create FlashMLA-ETAP, achieving enhanced inference efficiency.

\subsection{Efficient Transpose Attention Pipeline (ETAP)}
\subsubsection{Original MLA Computation Mode in Inference}
During the inference phase, MLA follows a standard computation pipeline for each head, derived from the attention mechanism. Given the query, key, and value matrices $\mathbf{Q}, \mathbf{K}, \mathbf{V} \in \mathbb{R}^{N \times d}$, where $N$ is the sequence length and $d$ is the head dimension, the attention output $\mathbf{O}$ is computed as:

\[
\mathbf{S} = \mathbf{Q} \cdot \mathbf{K}^T \in \mathbb{R}^{N \times N}, \quad \mathbf{P} = \text{softmax}(\mathbf{S}) \in \mathbb{R}^{N \times N}, \quad \mathbf{O} = \mathbf{P} \cdot \mathbf{V} \in \mathbb{R}^{N \times d}.
\]

In practice, MLA applies low-rank joint compression to $\mathbf{K}$ and $\mathbf{V}$, reducing the KV cache size, which is critical for large models like DeepSeek-V3. However, during inference, especially in decoding phases, the query length is typically short (e.g., 1 or 2 tokens), while the KV cache context length can be extremely long (e.g., 1K to 32K tokens), exacerbating the computational disparity.

\subsubsection{Challenges with the Original Mode}
The original computation mode poses significant challenges on the NVIDIA H20 GPU, primarily due to the mismatch between the workload and the GPU's architectural constraints. First, the WGMMA instruction, optimized for Hopper GPUs, requires the \(M\)-dimension to be at least 64 for efficient execution. In decoding scenarios, the per-GPU head count, reduced to 16 due to insufficient memory necessitating head splitting of the DeepSeek-R1 671B model’s 128 heads across 8 GPUs, falls below this minimum, causing redundant padding and a compute utilization of only 25\% or lower. Second, the long KV context length results in large $\mathbf{K}$ and $\mathbf{V}$ matrices, increasing memory access latency and HBM bandwidth pressure on the H20, which lacks the abundant resources of high-end GPUs like the H100/H800. These inefficiencies manifest as prolonged inference times, with MLA accounting for approximately 30\% of a decoding forward pass in models like DeepSeek-V3 (e.g., BS=16, ContextLength=16K), severely impacting real-time performance in resource-constrained environments.

\subsubsection{ETAP: A Novel Computation Mode}
To address these challenges, we propose the Efficient Transpose Attention Pipeline (ETAP), a novel computation mode that leverages transposition to reduce redundant calculations. ETAP reconfigures the standard attention computation by transposing the operations, exploiting the asymmetry between query and KV lengths during inference. Specifically, instead of computing:

\[
\mathbf{S} = \mathbf{Q} \cdot \mathbf{K}^T, \quad \mathbf{P} = \text{softmax}(\mathbf{S}), \quad \mathbf{O} = \mathbf{P} \cdot \mathbf{V},
\]

ETAP computes:

\begin{align}
\mathbf{S}^T &= \mathbf{K} \cdot \mathbf{Q}^T \in \mathbb{R}^{N \times N_q}, \\
\mathbf{P}^T &= \text{softmax}(\mathbf{S}^T) \in \mathbb{R}^{N \times N_q}, \\
\mathbf{O} &= \mathbf{V}^T \cdot \mathbf{P}^T \in \mathbb{R}^{d \times N_q}, \\
\mathbf{O} &= \mathbf{O}^T,
\end{align}

where $N_q$ is the query length and $N$ is the KV context length. By treating the KV context length as the $M$-dimension and the number of query heads as the $N$-dimension in WGMMA operations, ETAP aligns the workload with the H20's architectural constraints, eliminating the need for redundant padding on the query dimension. The final transpose ($\mathbf{O} = \mathbf{O}^T$) is performed once, while the inner product computations benefit from repeated efficiency gains across the long context length.

\subsubsection{Theoretical Analysis of ETAP Benefits}
The primary benefit of ETAP lies in its ability to reduce padding overhead and improve compute utilization on the H20’s 148 TFLOPS FP16 capability. In the original mode, the $M$-dimension (query length) requires padding to meet WGMMA requirements when it is small, introducing a padding factor that scales inversely with query length. ETAP mitigates this by redefining the $M$-dimension as the KV context length, which is naturally large and requires no padding, while the $N$-dimension (query-related) is handled efficiently without additional constraints. This reorientation reduces the proportion of redundant computations, theoretically enhancing compute efficiency. The exact improvement depends on the relative sizes of the query length and KV context length, with greater benefits observed as the context length increases relative to the query length. This approach offers a scalable solution to the padding problem, optimizing the attention mechanism for mid-tier GPUs with constrained compute resources.

\subsection{FlashMLA-ETAP Algorithm}
\subsubsection{Integration into MLA Frameworks}
The simplicity and generality of ETAP make it a versatile computation mode that can be seamlessly integrated into various MLA frameworks, including FlashAttention-3, FlashInfer, and FlashMLA. By reorienting the attention computation through transposition, ETAP requires minimal changes to existing pipelines, primarily in the matrix multiplication and softmax stages, while preserving numerical stability and compatibility with low-rank compression techniques like those in MLA. This compatibility stems from the shared attention computation pattern across these frameworks—\(\mathbf{S} = \mathbf{Q} \cdot \mathbf{K}^T\), \(\mathbf{P} = \text{softmax}(\mathbf{S})\), \(\mathbf{O} = \mathbf{P} \cdot \mathbf{V}\)—which ETAP reconfigures into a transposed form that aligns with the H20’s WGMMA constraints. A detailed theoretical analysis of ETAP’s integration into FlashAttention-3 and FlashInfer, highlighting their architectural similarities and the minimal modifications required, is provided in the end of this section.

\subsubsection{FlashMLA-ETAP: Algorithm and Implementation}

FlashMLA-ETAP extends FlashMLA by incorporating the ETAP computation mode, tailoring the attention pipeline for the H20's 
hardware constraints. The core modification lies in the transposition of the attention computation, which we implement within FlashMLA's cooperative thread array (CTA) framework, leveraging producer-consumer synchronization and block-wise processing. Below, we present the pseudocode for FlashMLA-ETAP's forward pass, followed by a detailed explanation.

\begin{algorithm*}
\caption{\textsc{FlashMLA-ETAP} Forward Pass \textbf{with} Intra-Consumer Overlapping -- CTA View}
\begin{algorithmic}[1]
\Require Matrices $\mathbf{Q}_i \in \mathbb{R}^{B_r \times d}$ and $\mathbf{K}, \mathbf{V} \in \mathbb{R}^{N \times d}$ in HBM, key block size $B_c$ with $T_c = \left\lceil \frac{N}{B_c} \right\rceil$. $\mathbf{V} = [\mathbf{V_0}, \mathbf{V_1}]$, $\mathbf{V_0}, \mathbf{V_1} \in \mathbb{R}^{N/2 \times d}$

\State Initialize pipeline object to manage barrier synchronization with $s$-stage circular SMEM buffer.
\State On-chip, initialize $\mathbf{O}_i = \mathbf{0} \in \mathbb{R}^{d \times B_r}$ and $\ell_i, m_i = 0, (-\infty) \in \mathbb{R}^{B_r}$, $\mathbf{O}_i = [\mathbf{O}_{i0}, \mathbf{O}_{i1}]$, $\mathbf{O}_{i0}, \mathbf{O}_{i1} \in \mathbb{R}^{d/2 \times B_r}$

\If{in consumer warpgroup0}
  \State Reallocate predetermined number of registers as function of number of consumer warpgroup0.
  \State Wait for $\mathbf{Q}_i$ to be loaded in shared memory.
  \For{$0 \leq j < T_c$}
    \State Syncthreads. Wait for $\mathbf{K}_j$ to be loaded in shared memory.
    \State Compute $\mathbf{S}_i^{(j)} = \mathbf{K}_j \mathbf{Q}_i^\top$ (SS-GEMM). Commit and wait.
    \State Store $m_i^{\text{old}} = m_i$ and compute $m_i = \max(m_i^{\text{old}}, \text{rowmax}(\mathbf{S}_i^{(j)}))$.
    \State Compute $\widetilde{\mathbf{P}}_i^{(j)} = \exp(\mathbf{S}_i^{(j)} - m_i)$ and $\ell_i = \exp(m_i^{\text{old}} - m_i)\ell_i + \text{colsum}(\widetilde{\mathbf{P}}_i^{(j)})$.
    \State Wait for $\mathbf{V}_j$ to be loaded in shared memory.
    \State Compute $\mathbf{R}_i = \operatorname{diag}(\exp(m_i^{\text{old}} - m_i))^{-1}$
    \State Store $\mathbf{R}_i$ to shared memory
    \State Compute $\mathbf{O}_{i0} = \mathbf{R}_i \mathbf{O}_{i0} + \mathbf{V}_{j0}^\top \widetilde{\mathbf{P}}_i^{(j)}$ (SS-GEMM). Commit and wait.
    \State Release the $(j \bmod s)$-th stage of the buffer for the producer.
  \EndFor
\Else{in producer warpgroup1}
  \State Reallocate predetermined number of registers as function of number of producer warpgroup1.
  \State Issue load $\mathbf{Q}_i$ from HBM to shared memory.
  \State Issue load $\mathbf{K}_0$ from HBM to shared memory.
  \For{$1 \leq j < T_c$}
    \State CopyAsyncWait and Syncthreads
    \State Issue loads of $\mathbf{K}_j, \mathbf{V}_j$ from HBM to shared memory at the $(j \bmod s)$-th stage of the buffer.
    \State NamedBarrier.Sync warpgroup0 and warpgroup1
    \State Copy $\mathbf{R}_i$ from shared memory.
    \State Compute $\mathbf{O}_{i1} = \mathbf{R}_i \mathbf{O}_{i1} + \mathbf{V}_{j1}^\top \widetilde{\mathbf{P}}_i^{(j)}$ (SS-GEMM). Commit and wait.
  \EndFor
\EndIf
\State Compute $\mathbf{O}_i = \operatorname{diag}(\ell_i)^{-1} \mathbf{O}_i$ and $L_i = m_i + \log(\ell_i)$.
\State Transpose $\mathbf{O}_i = \mathbf{O}_i^\top$.
\State Write $\mathbf{O}_i$ and $L_i$ to HBM as the $i$-th block of $\mathbf{O}$ and $L$.
\end{algorithmic}
\end{algorithm*}

\subsubsection{Explanation of FlashMLA-ETAP Pseudocode}
The FlashMLA-ETAP algorithm operates within a CTA, utilizing producer-consumer synchronization to manage data movement and computation. The pipeline is initialized with a circular shared memory (SMEM) buffer to handle block-wise processing (Line 1). The output matrix $\mathbf{O}_i$ and softmax statistics ($\ell_i, m_i$) are initialized on-chip, with $\mathbf{O}_i$ split into two segments ($\mathbf{O}_{i0}, \mathbf{O}_{i1}$) for intra-consumer overlapping (Line 2). The computation is divided between two warpgroups: consumer warpgroup0 and producer warpgroup1.

\begin{itemize}
    \item \textbf{Consumer Warpgroup0 (Lines 4--14)}: This warpgroup computes the attention scores using the transposed form $\mathbf{S}_i^{(j)} = \mathbf{K}_j \mathbf{Q}_i^\top$, reflecting ETAP's approach (Line 7). It calculates the softmax statistics (Lines 8--9) and updates the first segment of the output $\mathbf{O}_{i0}$ (Lines 10--12), leveraging shared memory to store intermediate results ($\mathbf{R}_i$) for synchronization.
    \item \textbf{Producer Warpgroup1 (Lines 15--23)}: This warpgroup handles data loading from HBM to SMEM (Lines 16--19), synchronizes with the consumer via named barriers (Line 20), and computes the second segment $\mathbf{O}_{i1}$ using the shared $\mathbf{R}_i$ (Lines 21--22). The transposition inherent in ETAP ensures efficient dimension alignment with WGMMA instructions.
    \item \textbf{Epilogue (Lines 24--26)}: The final steps rescale the output, apply the ETAP-required transpose ($\mathbf{O}_i = \mathbf{O}_i^\top$), and write results to HBM.
\end{itemize}

This implementation minimizes padding overhead by aligning the KV context length with the \(M\)-dimension, achieving significant performance improvements on the NVIDIA H20 while maintaining compatibility with FlashMLA’s low-rank compression and inference optimizations.

\subsubsection{Theoretical Analysis of ETAP’s Integration into FlashAttention-3 and FlashInfer}
To substantiate the claim that ETAP can be seamlessly integrated into other attention frameworks such as FlashAttention-3 \cite{shah2024} and FlashInfer \cite{ye2025}, we provide a theoretical analysis of its compatibility with their computational pipelines, focusing on the structural similarities and differences that influence integration feasibility. Both FlashAttention-3 and FlashInfer, like FlashMLA, are designed to optimize attention mechanisms on modern GPUs, but they differ in their implementation strategies and target hardware optimizations, necessitating a careful examination of how ETAP’s transposition-based approach aligns with their architectures.

FlashAttention-3, an evolution of the FlashAttention series, employs I/O-aware tiling, online softmax computation, and asynchronous execution to minimize high-bandwidth memory (HBM) accesses, leveraging advanced GPU features like Tensor Memory Accelerator (TMA). Its core pipeline computes attention scores \(\mathbf{S} = \mathbf{Q} \cdot \mathbf{K}^T\), applies softmax to obtain probabilities \(\mathbf{P} = \text{softmax}(\mathbf{S})\), and computes the output \(\mathbf{O} = \mathbf{P} \cdot \mathbf{V}\), similar to the standard MLA computation in FlashMLA. However, FlashAttention-3’s tiling strategy partitions the computation into blocks to optimize memory access, a process that can introduce padding overhead on mid-tier GPUs like the H20 when the query length is small (e.g., 1 or 2 tokens) compared to the KV context length (e.g., 16K tokens). ETAP’s transposition approach, which reconfigures the computation as \(\mathbf{S}^T = \mathbf{K} \cdot \mathbf{Q}^T\), \(\mathbf{P}^T = \text{softmax}(\mathbf{S}^T)\), and \(\mathbf{O} = (\mathbf{V}^T \cdot \mathbf{P}^T)^T\), directly addresses this inefficiency by aligning the KV context length with the \(M\)-dimension in WGMMA operations, reducing padding overhead. Integrating ETAP into FlashAttention-3 requires modifying its matrix multiplication and softmax stages to adopt this transposed computation flow. Since FlashAttention-3 already employs block-wise processing, the transposition can be applied at the block level, preserving its I/O-aware tiling strategy. The primary adjustment involves redefining the block dimensions to prioritize the KV context length as the \(M\)-dimension, which can be achieved with minimal changes to the existing pipeline, as the softmax and output computation stages remain functionally equivalent after transposition. Additionally, FlashAttention-3’s use of TMA for asynchronous memory transfers is unaffected by ETAP’s modifications, ensuring compatibility with its high-end GPU optimizations while extending its efficiency to mid-tier GPUs.

FlashInfer, another optimized attention framework, focuses on efficient inference for long-context scenarios, often employing kernel fusion and custom CUDA implementations to reduce overheads. Unlike FlashMLA, which incorporates low-rank joint compression for KV caches, FlashInfer typically operates on full attention matrices but shares the same fundamental computation pattern: \(\mathbf{S} = \mathbf{Q} \cdot \mathbf{K}^T\), \(\mathbf{P} = \text{softmax}(\mathbf{S})\), and \(\mathbf{O} = \mathbf{P} \cdot \mathbf{V}\). FlashInfer’s design emphasizes latency reduction through fused kernels, which can exacerbate padding inefficiencies on the H20 when the query length is small, similar to FlashAttention-3. ETAP’s transposition strategy can be integrated into FlashInfer by modifying its fused kernels to compute \(\mathbf{S}^T = \mathbf{K} \cdot \mathbf{Q}^T\) and subsequent steps in the transposed form. This adjustment requires refactoring the kernel to handle the transposed dimensions, but the core computation—matrix multiplication followed by softmax—remains structurally identical, ensuring that FlashInfer’s kernel fusion benefits are preserved. Moreover, since FlashInfer does not rely on low-rank compression, ETAP’s compatibility with such techniques (as demonstrated in FlashMLA) is not a concern, simplifying the integration process. The transposition can be implemented as a preprocessing step within the kernel, aligning the workload with the H20’s WGMMA constraints without altering FlashInfer’s overall optimization strategy.

Theoretically, ETAP’s integration into both frameworks benefits from its focus on a universal challenge in attention mechanisms: the padding overhead in decoding scenarios with short queries and long KV contexts. This issue is not specific to FlashMLA but affects any framework operating on mid-tier GPUs with similar architectural constraints. By reorienting the computation to minimize padding, ETAP enhances compute utilization across frameworks, with the primary integration cost being the adjustment of matrix multiplication and softmax stages to accommodate the transposed computation. The numerical stability of ETAP, as validated in our experiments with FlashMLA-ETAP, further supports its applicability, as the transposition does not introduce significant floating-point errors, a critical consideration for frameworks like FlashAttention-3 that support low-precision formats such as FP8. Additionally, ETAP’s reliance on WGMMA instructions, while optimized for the H20, is compatible with other Hopper-based GPUs, suggesting that its benefits could extend to mid-tier platforms beyond the H20 with similar instruction sets. This analysis underscores ETAP’s design as a generalizable solution, capable of enhancing the efficiency of diverse attention frameworks on mid-tier GPUs, thereby supporting its potential for broader adoption in hardware-aware inference optimization.

\section{Experimental Evaluation}
\subsection{Experimental Setup}

To evaluate the performance of FlashMLA-ETAP, we conducted experiments comparing its inference throughput against FlashAttention-3, FlashInfer, and FlashMLA using the DeepSeek-R1 model on an NVIDIA H20 GPU. The H20, a mid-tier GPU based on the NVIDIA Hopper architecture, is equipped with 96GB of HBM3 memory, offering a memory bandwidth of 4.0TB/s. Its BF16/FP16 compute capability reaches 148 TFLOPS, making it a suitable platform for assessing hardware-specific optimizations like ETAP in resource-constrained environments. Its architectural features, such as WGMMA instruction padding requirements, highlight the need for tailored solutions to maximize compute utilization. Experiments simulate auto-regressive decoding by generating one token per forward pass, utilizing 16 attention heads with a head dimension of 576, consistent with the DeepSeek-R1 configuration. We measured inference performance in TFLOPS/s across sequence lengths of 512, 1K, 2K, 4K, 8K, 16K, 32K, and 64K, covering a range of practical scenarios from short-context to long-context tasks. Experiments were performed with batch sizes of 16 and 32, using FP16 precision to align with the H20’s capabilities, with each test repeated five times to compute the average TFLOPS/s, mitigating runtime variability. The setup excluded extraneous optimizations to focus on the core attention mechanisms.

\begin{figure}[ht]
    \centering
    \includegraphics[width=0.85\textwidth,height=0.60\textheight]{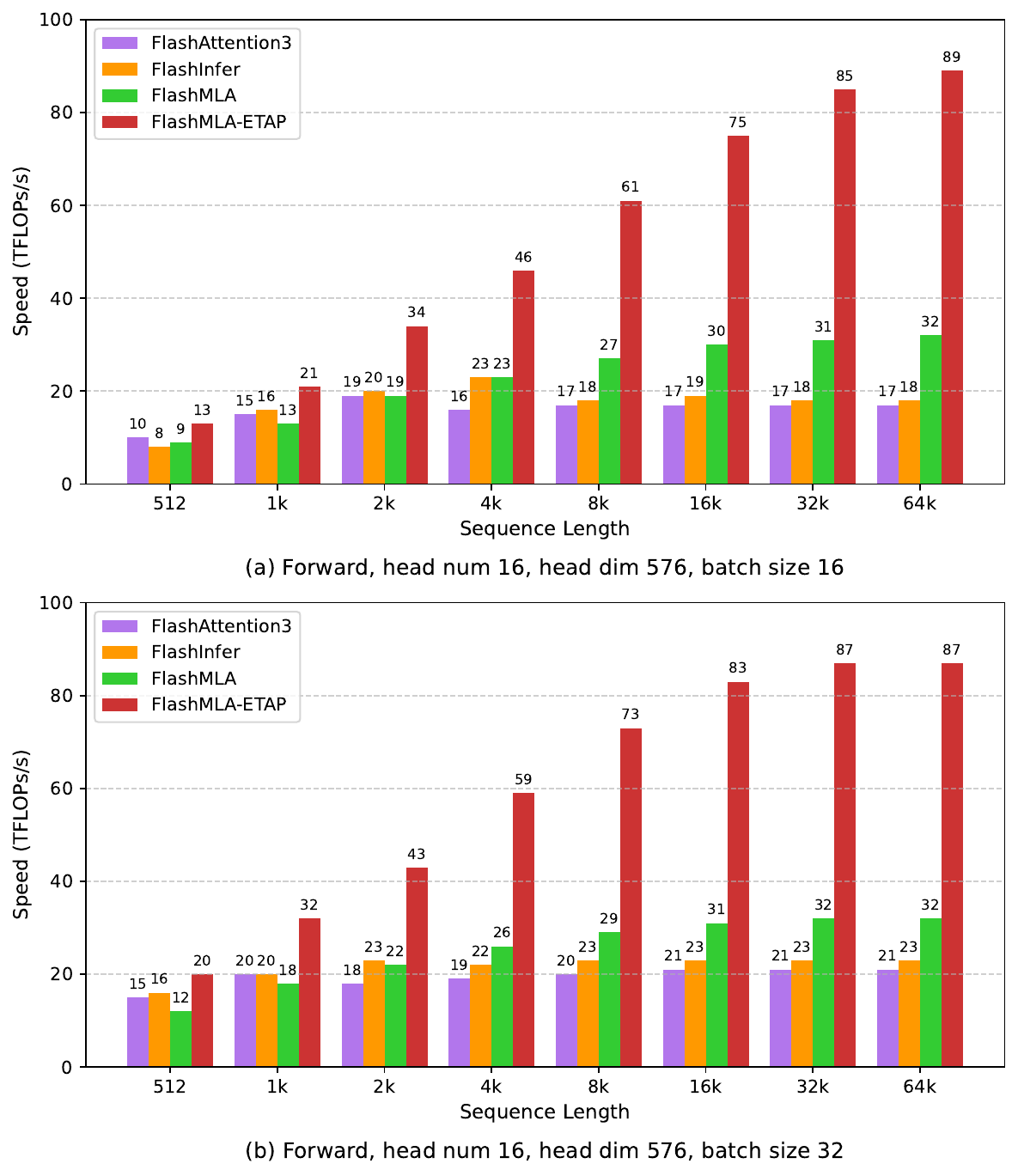}
    \caption{Inference performance comparison (TFLOPS/s) on the NVIDIA H20 GPU using the DeepSeek-R1 model with head num 16 and head dimension 576. (a) Batch size 16. (b) Batch size 32.}
    \label{fig:attention_speed_comparison}
\end{figure}

\subsection{Results and Analysis}

Figure~\ref{fig:attention_speed_comparison} presents the inference performance of FlashMLA-ETAP, FlashAttention-3, FlashInfer, and FlashMLA across the tested sequence lengths for batch sizes 16 (subfigure a) and 32 (subfigure b), visualized as bar charts. The data, derived from experiments on the DeepSeek-R1 model, highlights the throughput in TFLOPS/s for each framework.

FlashMLA-ETAP demonstrates superior performance across all sequence lengths and batch sizes. For batch size 16, it achieves a peak of 89 TFLOPS/s at 64K, a 2.78x improvement over FlashMLA's 32 TFLOPS/s at the same length, with the speedup growing from 1.44x (13/9) at 512 to 2.78x at 64K. This trend, evident in Figure~\ref{fig:attention_speed_comparison}(a), reflects ETAP’s effectiveness in reducing padding overhead as sequence length increases, aligning the KV context length with the \(M\)-dimension in WGMMA operations. Compared to FlashAttention-3 (17 TFLOPS/s at 64K) and FlashInfer (18 TFLOPS/s at 64K), FlashMLA-ETAP offers 5.24x and 4.94x improvements, respectively, underscoring its optimization for the H20’s 148 TFLOPS FP16 constraints. For batch size 32, FlashMLA-ETAP peaks at 87 TFLOPS/s at 32K and 64K, a 2.72x gain over FlashMLA's 32 TFLOPS/s, with 4.14x and 3.78x advantages over FlashAttention-3 (21 TFLOPS/s) and FlashInfer (23 TFLOPS/s) at 64K, as shown in Figure~\ref{fig:attention_speed_comparison}(b).

The performance advantage is most pronounced at longer sequence lengths (16K to 64K), where FlashMLA-ETAP reaches 61 to 89 TFLOPS/s (batch size 16) and 73 to 87 TFLOPS/s (batch size 32), compared to FlashMLA's 27 to 32 TFLOPS/s and 29 to 32 TFLOPS/s. FlashAttention-3 and FlashInfer exhibit flatter profiles (10-21 TFLOPS/s and 8-23 TFLOPS/s), indicating their optimization for high-end GPUs, which underperforms on the H20 due to its compute constraints. The slight performance boost with batch size 32 (e.g., 87 vs. 89 TFLOPS/s at 64K) suggests effective parallelization, though gains plateau beyond 32K, likely due to compute saturation rather than memory bandwidth.

\subsection{Numerical Error Validation}

To ensure the numerical stability of FlashMLA-ETAP, we conducted a validation experiment comparing its RMSE against FlashAttention-3, following a methodology similar to that used in the FlashAttention-3 paper. We computed the RMSE between the FP16 outputs of each framework and a double-precision (FP64) reference implementation of the attention mechanism, using the DeepSeek-R1 model across a representative set of sequence lengths and batch sizes on the H20 GPU. Table~\ref{tab:numerical_error} summarizes the results.

\begin{table}[ht]
    \centering
    \caption{RMSE comparison between FlashAttention-3 and FlashMLA-ETAP in FP16 precision.}
    \setlength{\tabcolsep}{1cm}
    \renewcommand{\arraystretch}{1.2}
    \begin{tabular}{lcc}
        \toprule
        \textbf{Framework} & \textbf{RMSE} \\
        \midrule
        FlashAttention-3 & \(1.9 \times 10^{-4}\) \\
        FlashMLA-ETAP    & \(1.25 \times 10^{-5}\) \\
        \bottomrule
    \end{tabular}
    \label{tab:numerical_error}
\end{table}

FlashMLA-ETAP exhibits a significantly lower RMSE (\(1.25 \times 10^{-5}\)) compared to FlashAttention-3 (\(1.9 \times 10^{-4}\)), representing a reduction by a factor of approximately 15.2x. This improvement indicates that FlashMLA-ETAP maintains higher numerical accuracy despite its optimizations, such as the transposition in ETAP and the integration with FlashMLA's low-rank compression. The reduced error suggests that the reconfigured computation mode effectively preserves the integrity of the attention mechanism, even under the H20's FP16 compute constraints (148 TFLOPS), making it a robust choice for inference tasks requiring precision.

\subsection{Discussion}

The experimental results affirm the effectiveness of FlashMLA-ETAP, with its 2.78x speedup over FlashMLA at 64K (batch size 16) validating the Efficient Transpose Attention Pipeline (ETAP) as a tailored optimization for the H20’s 148 TFLOPS FP16 compute capability. The increasing performance gap with sequence length, as shown in Figure~\ref{fig:attention_speed_comparison}, highlights ETAP’s ability to mitigate padding overhead, a key bottleneck on mid-tier hardware. The significant improvements over FlashAttention-3 (5.24x) and FlashInfer (4.94x) at 64K demonstrate that ETAP addresses the H20’s specific constraints more effectively than frameworks optimized for high-end GPUs.

The scalability across batch sizes 16 and 32, with modest gains at larger batches (e.g., 87 vs. 89 TFLOPS/s at 64K), indicates robust handling of parallel workloads, suitable for diverse inference scenarios. The plateau beyond 32K suggests a potential compute saturation limit, warranting further investigation. The numerical error validation reveals FlashMLA-ETAP’s superior accuracy (RMSE \(1.25 \times 10^{-5}\)) compared to FlashAttention-3 (\(1.9 \times 10^{-4}\)), reinforcing its reliability for precision-sensitive applications. Future work could focus on empirically validating ETAP’s integration into FlashAttention-3 and FlashInfer, as supported by the theoretical analysis in Section 3, and optimizing it for other Hopper-based mid-tier GPUs to broaden its applicability.

\section{Conclusion, Limitations, Future Work}
This paper introduces FlashMLA-ETAP, a novel framework that significantly enhances Multi-Head Latent Attention (MLA) inference on the NVIDIA H20 GPU, constrained by its 148 TFLOPS FP16 compute capability. By proposing the Efficient Transpose Attention Pipeline (ETAP), which reconfigures attention computation to reduce padding overhead through transposition, FlashMLA-ETAP achieves a 2.78x speedup over FlashMLA at 64K sequence length (batch size 16), outperforming FlashAttention-3 by 5.24x and FlashInfer by 4.94x, as demonstrated in our experiments. ETAP’s design also proves theoretically compatible with FlashAttention-3 and FlashInfer, highlighting its potential for broader adoption in hardware-aware inference optimization. Additionally, FlashMLA-ETAP exhibits a 15.2x lower RMSE (\(1.25 \times 10^{-5}\)) compared to FlashAttention-3, ensuring numerical stability in long-context scenarios. However, the evaluation is limited to the H20, autoregressive decoding with one token per forward pass, and specific configurations (16 heads, 576 head dimension, up to 64K sequences), potentially restricting generalizability across diverse GPUs, inference paradigms, or larger models. Future work should focus on empirically validating ETAP’s integration into FlashAttention-3 and FlashInfer, optimizing for other Hopper-based mid-tier GPUs, and exploring broader sequence lengths and configurations to enhance scalability and applicability.

%
%
%
\bibliographystyle{splncs04}
\bibliography{mybibliography}

@inproceedings{vaswani2017,
 author = {Vaswani, Ashish and Shazeer, Noam and Parmar, Niki and Uszkoreit, Jakob and others},
 booktitle = {Advances in Neural Information Processing Systems},
 editor = {I. Guyon and U. Von Luxburg and S. Bengio and H. Wallach and R. Fergus and S. Vishwanathan and R. Garnett},
 pages = {},
 publisher = {Curran Associates, Inc.},
 title = {Attention is All you Need},
 volume = {30},
 year = {2017}
}

@misc{deepseek-ai2024,
      title={DeepSeek-V2: A Strong, Economical, and Efficient Mixture-of-Experts Language Model}, 
      author={DeepSeek-AI and Aixin Liu and Bei Feng and Bin Wang and others},
      year={2024},
      eprint={2405.04434},
      archivePrefix={arXiv},
      primaryClass={cs.CL},
      url={https://arxiv.org/abs/2405.04434}, 
}

@inproceedings{dao2022,
 author = {Dao, Tri and Fu, Dan and Ermon, Stefano and Rudra, Atri and others},
 booktitle = {Advances in Neural Information Processing Systems},
 editor = {S. Koyejo and S. Mohamed and A. Agarwal and D. Belgrave and K. Cho and A. Oh},
 pages = {16344--16359},
 publisher = {Curran Associates, Inc.},
 title = {FlashAttention: Fast and Memory-Efficient Exact Attention with IO-Awareness},
 volume = {35},
 year = {2022}
}

@misc{dao2023,
      title={FlashAttention-2: Faster Attention with Better Parallelism and Work Partitioning}, 
      author={Tri Dao},
      year={2023},
      eprint={2307.08691},
      archivePrefix={arXiv},
      primaryClass={cs.LG},
      url={https://arxiv.org/abs/2307.08691}, 
}

@misc{shah2024,
      title={FlashAttention-3: Fast and Accurate Attention with Asynchrony and Low-precision}, 
      author={Jay Shah and Ganesh Bikshandi and Ying Zhang and Vijay Thakkar and Pradeep Ramani and Tri Dao},
      year={2024},
      eprint={2407.08608},
      archivePrefix={arXiv},
      primaryClass={cs.LG},
      url={https://arxiv.org/abs/2407.08608}, 
}

@misc{ye2025,
      title={FlashInfer: Efficient and Customizable Attention Engine for LLM Inference Serving}, 
      author={Zihao Ye and Lequn Chen and Ruihang Lai and Wuwei Lin and Yineng Zhang and Stephanie Wang and Tianqi Chen and Baris Kasikci and Vinod Grover and Arvind Krishnamurthy and Luis Ceze},
      year={2025},
      eprint={2501.01005},
      archivePrefix={arXiv},
      primaryClass={cs.DC},
      url={https://arxiv.org/abs/2501.01005}, 
}

@misc{flashmla2025,
      title={FlashMLA: Efficient MLA decoding kernels},
      author={Jiashi Li, Shengyu Liu},
      year={2025},
      publisher = {GitHub},
      howpublished = {\url{https://github.com/deepseek-ai/FlashMLA}},
}

@misc{sglang2024,
      title={SGLang:  a fast serving framework for large language models and vision language models},
      author={sgl},
      year={2024},
      publisher = {GitHub},
      howpublished = {\url{https://github.com/sgl-project/sglang}},
}

@misc{deepseek-ai2025,
      title={DeepSeek-V3 Technical Report}, 
      author={DeepSeek-AI and Aixin Liu and Bei Feng and Bing Xue and others},
      year={2025},
      eprint={2412.19437},
      archivePrefix={arXiv},
      primaryClass={cs.CL},
      url={https://arxiv.org/abs/2412.19437}, 
}

@misc{deepseek-ai2025r1,
      title={DeepSeek-R1: Incentivizing Reasoning Capability in LLMs via Reinforcement Learning}, 
      author={DeepSeek-AI and Daya Guo and Dejian Yang and Haowei Zhang and others},
      year={2025},
      eprint={2501.12948},
      archivePrefix={arXiv},
      primaryClass={cs.CL},
      url={https://arxiv.org/abs/2501.12948}, 
}

@inproceedings{devlin2019,
    title = "{BERT}: Pre-training of Deep Bidirectional Transformers for Language Understanding",
    author = "Devlin, Jacob  and
      Chang, Ming-Wei  and
      Lee, Kenton  and
      Toutanova, Kristina",
    editor = "Burstein, Jill  and
      Doran, Christy  and
      Solorio, Thamar",
    booktitle = "Proceedings of the 2019 Conference of the North {A}merican Chapter of the Association for Computational Linguistics: Human Language Technologies, Volume 1 (Long and Short Papers)",
    month = jun,
    year = "2019",
    address = "Minneapolis, Minnesota",
    publisher = "Association for Computational Linguistics",
    url = "https://aclanthology.org/N19-1423/",
    doi = "10.18653/v1/N19-1423",
    pages = "4171--4186",
}

@misc{alexey2020,
      title={An Image is Worth 16x16 Words: Transformers for Image Recognition at Scale}, 
      author={Alexey Dosovitskiy and Lucas Beyer and Alexander Kolesnikov and Dirk Weissenborn and others},
      year={2021},
      eprint={2010.11929},
      archivePrefix={arXiv},
      primaryClass={cs.CV},
      url={https://arxiv.org/abs/2010.11929}, 
}

@InProceedings{radford2021,
  title = 	 {Learning Transferable Visual Models From Natural Language Supervision},
  author =       {Radford, Alec and Kim, Jong Wook and Hallacy, Chris and Ramesh, Aditya and others},
  booktitle = 	 {Proceedings of the 38th International Conference on Machine Learning},
  pages = 	 {8748--8763},
  year = 	 {2021},
  editor = 	 {Meila, Marina and Zhang, Tong},
  volume = 	 {139},
  series = 	 {Proceedings of Machine Learning Research},
  month = 	 {18--24 Jul},
  publisher =    {PMLR},
}

@misc{ji2025,
      title={Towards Economical Inference: Enabling DeepSeek's Multi-Head Latent Attention in Any Transformer-based LLMs}, 
      author={Tao Ji and Bin Guo and Yuanbin Wu and Qipeng Guo and others},
      year={2025},
      eprint={2502.14837},
      archivePrefix={arXiv},
      primaryClass={cs.CL},
      url={https://arxiv.org/abs/2502.14837}, 
}

@misc{huang2024,
      title={Advancing Transformer Architecture in Long-Context Large Language Models: A Comprehensive Survey}, 
      author={Yunpeng Huang and Jingwei Xu and Junyu Lai and Zixu Jiang and others},
      year={2024},
      eprint={2311.12351},
      archivePrefix={arXiv},
      primaryClass={cs.CL},
      url={https://arxiv.org/abs/2311.12351}, 
}

@misc{child2019,
      title={Generating Long Sequences with Sparse Transformers}, 
      author={Rewon Child and Scott Gray and Alec Radford and Ilya Sutskever},
      year={2019},
      eprint={1904.10509},
      archivePrefix={arXiv},
      primaryClass={cs.LG},
      url={https://arxiv.org/abs/1904.10509}, 
}

@misc{beltagy2020,
      title={Longformer: The Long-Document Transformer}, 
      author={Iz Beltagy and Matthew E. Peters and Arman Cohan},
      year={2020},
      eprint={2004.05150},
      archivePrefix={arXiv},
      primaryClass={cs.CL},
      url={https://arxiv.org/abs/2004.05150}, 
}

@misc{liu2023,
      title={Ring Attention with Blockwise Transformers for Near-Infinite Context}, 
      author={Hao Liu and Matei Zaharia and Pieter Abbeel},
      year={2023},
      eprint={2310.01889},
      archivePrefix={arXiv},
      primaryClass={cs.CL},
      url={https://arxiv.org/abs/2310.01889}, 
}

@misc{gu2024,
      title={Mamba: Linear-Time Sequence Modeling with Selective State Spaces}, 
      author={Albert Gu and Tri Dao},
      year={2024},
      eprint={2312.00752},
      archivePrefix={arXiv},
      primaryClass={cs.LG},
      url={https://arxiv.org/abs/2312.00752}, 
}

@misc{peng2023,
      title={RWKV: Reinventing RNNs for the Transformer Era}, 
      author={Bo Peng and Eric Alcaide and Quentin Anthony and Alon Albalak and others},
      year={2023},
      eprint={2305.13048},
      archivePrefix={arXiv},
      primaryClass={cs.CL},
      url={https://arxiv.org/abs/2305.13048}, 
}

@inproceedings{kwon2023,
author = {Kwon, Woosuk and Li, Zhuohan and Zhuang, Siyuan and Sheng, Ying and Zheng, Lianmin and Yu, Cody Hao and Gonzalez, Joseph and Zhang, Hao and Stoica, Ion},
title = {Efficient Memory Management for Large Language Model Serving with PagedAttention},
year = {2023},
isbn = {9798400702297},
publisher = {Association for Computing Machinery},
address = {New York, NY, USA},
url = {https://doi.org/10.1145/3600006.3613165},
doi = {10.1145/3600006.3613165},
booktitle = {Proceedings of the 29th Symposium on Operating Systems Principles},
pages = {611–626},
numpages = {16},
location = {Koblenz, Germany},
series = {SOSP '23}
}

\end{document}